\documentclass[10pt,american,english,aps,showkeys,showpacs,nofootinbib]{revtex4-1}
\usepackage[T1]{fontenc}
\usepackage{textcomp}
\usepackage[utf8]{inputenc}
\usepackage{color}
\usepackage{babel}
\usepackage{amsmath}
\usepackage{amssymb}
\usepackage{stackrel}
\usepackage{graphicx}
\usepackage{geometry}
\usepackage[bookmarks=false,
 breaklinks=false,pdfborder={0 0 1},backref=section,colorlinks=false]
 {hyperref}

\makeatletter
\usepackage{babel}
\usepackage{todonotes}
\usepackage[bottom]{footmisc}
\usepackage{graphicx}
\usepackage{dcolumn}
\usepackage{bm}
\usepackage{todonotes}
\makeatother

\begin{document}
\title{Quantum Dynamics of Scalar Particles in a Spinning Cosmic String Background
with Topological Defects: A Feshbach-Villars Formalism Perspective}
\author{Sarra Garah}
\email{sarra.garah@univ-tebessa.dz ; gareh64@gmail.com}

\affiliation{Laboratory of theoretical and applied Physics~\\
 Echahid Cheikh Larbi Tebessi University, Algeria}
\author{Abdelmalek Boumali}
\email{boumali.abdelmalek@gmail.com}

\affiliation{Laboratory of theoretical and applied Physics~\\
 Echahid Cheikh Larbi Tebessi University, Algeria}
\date{\today}
\selectlanguage{american}%
\begin{abstract}
We study the relativistic quantum dynamics of spin-0 particles in
the spacetime of a spinning\} cosmic string that carries both spacelike
disclination (conical deficit $\alpha$) and screw-dislocation (torsion
$J_{z}$) together with frame dragging ($J_{t}$). Using the Feshbach--Villars
(FV) reformulation of the Klein--Gordon equation, we obtain a first-order
Hamiltonian with a positive-definite density, enabling a clean probabilistic
interpretation for bosons in curved/topologically nontrivial backgrounds.
In the weak-field regime (retaining terms $\mathcal{O}(G)$ and discarding
the $\mathcal{O}(G^{2})$ contribution that would otherwise lead to
(double)-confluent Heun behavior), separation of variables in a finite
cylinder of radius $R_{0}$ yields a Bessel radial equation with an
effective index $\nu(\alpha,J_{t},J_{z};E,k)$ that mixes rotation
and torsion. The hard-wall condition $J_{\nu}(\kappa R_{0})=0$ quantizes
the spectrum,
\[
E_{n}^{2}=m^{2}+k^{2}+\big(j_{\nu,n}/R_{0}\big)^{2},
\]
Working in the stationary positive-energy sector, we derive closed-form
normalized eigenfunctions and FV densities, and we evaluate information-theoretic
indicators (Fisher information and Shannon entropy) directly from
the FV probability density. We find that increased effective confinement
(via geometry/torsion) enhances Fisher information and reduces position-space
Shannon entropy, quantitatively linking defect parameters to localization/complexity.
The FV framework thus provides a robust, computationally transparent
route to spectroscopy and information measures for scalar particles
in rotating/torsional string backgrounds, and it smoothly reproduces
the pure-rotation, pure-torsion, and flat-spacetime limits.
\end{abstract}
\maketitle
\selectlanguage{english}%

\section{Introduction }

Topological defects such as cosmic strings, monopoles, and domain
walls are expected outcomes of symmetry-breaking mechanisms in the
early universe and various condensed matter analogs. Among these,
cosmic strings---idealized as infinitely thin line defects---stand
out for their capacity to produce observable gravitational effects
while leaving spacetime locally flat except along their core. When
these strings possess intrinsic angular momentum or torsional features,
the resulting geometry is no longer trivial: it exhibits frame dragging
and screw dislocations, making it an ideal candidate for studying
the quantum dynamics of particles in spacetimes with both curvature
and torsion. These defects introduce nontrivial topological and geometrical
structures, such as conical singularities and Burgers vectors, that
influence the propagation of matter and fields. Thus, investigating
quantum fields in such backgrounds helps bridge the gap between general
relativity and quantum mechanics, and offers insights into the behavior
of matter in extreme gravitational regimes \cite{Vilenkin_1981,Vilenkin_1994,Katanaev_1999}.

The impact of defects has been studied in numerous research papers
addressing this topic. Geusa de A. Marques and Valdir B. Bezerra studied
a hydrogen atom in the background spacetimes generated by an infinitely
thin cosmic string and by a pointlike global monopole \cite{Marques_2002,Marques_2005}.
The Spin-0 oscillator field under a magnetic field in a cosmic string
spacetime was treated in \cite{Boumali_2014} . Generalized Dirac
oscillator under an external magnetic field in cosmic dislocation
spacetime in \cite{Chen_2020}. The two-dimensional Kemmer oscillator
under the influence of the gravitational field produced by cosmic
string spacetime and in the presence of a uniform magnetic field,
as well as without a magnetic field, was investigated in \cite{Messai_2015}.
Exact solutions of a two-dimensional Duffin--Kemmer--Petiau oscillator
subject to a Coulomb potential in the gravitational field of a cosmic
string in \cite{Boumali_2017}. Rotating effects on relativistic quantum
systems have been investigated in the background of the cosmic string
spacetime in several works \cite{Bakke_2009,Bakke_2010,Bakke_2013,Garcia_2017,Mota_2017}.
The influence of dislocation associated with the torsion of the manifold
have been widely investigated in the literature \cite{Garcia_2017,Castro_2016,Jaszek_2001}.
G. de A. Marques et al. have been analyzed quantum scattering of an
electron by a topological defect called dispiration with an externally
applied magnetic field \cite{Marques_2002,Marques_2005}. The relativistic
dynamics of a neutral particle with a magnetic dipole moment interacting
with an external electric field were investigated by Bakke et al.
\cite{Bakke_2010}, who studied the relativistic and non-relativistic
quantum dynamics of a neutral particle with a permanent magnetic dipole
moment interacting with two distinct field configurations in a cosmic
string spacetime. In the standard analyses of multiparticle scattering,
the incident beam is typically modeled as a plane wave. This approach
reveals that the scattering cross-section scales directly with the
Fourier transform of the correlation function that characterizes the
density fluctuations \cite{VanHove_1954}(for more details see Ref.
\cite{Messai_2025}).

Spin-0 particles are traditionally described by the Klein--Gordon
(KG) equation, which is relativistically covariant but second-order
in time. While mathematically consistent, the KG equation suffers
from interpretational difficulties when applied to single-particle
quantum mechanics. Chief among these is the fact that its probability
density, derived from the time component of a conserved current, is
not positive-definite. This precludes a straightforward probabilistic
interpretation of the wavefunction and leads to ambiguities in the
physical meaning of negative energy solutions. In curved spacetimes
or those with topological defects, these issues are further exacerbated,
limiting the utility of the KG framework in describing quantum phenomena
with geometric complexity\cite{Dirac_1981,Greiner_2000,Sakurai_1967}
.

To resolve these difficulties, the Feshbach--Villars (FV) transformation
offers an elegant and physically meaningful reformulation \cite{Feshbach_1958}.
It recasts the second-order KG equation into a first-order Schrödinger-like
form by decomposing the scalar field into two components, $\phi$
and $\chi$, which respectively represent the particle and antiparticle
sectors. This yields a two-component wavefunction$\Phi=(\phi,\chi)^{T}$
governed by a Hamiltonian with Pauli-type structure involving the
usual Pauli matrices $\tau_{i}$. As a result, the FV formalism enables
a consistent first-order temporal evolution and a positive-definite,
conserved probability density given by $\rho_{FVO}=\Phi^{\dagger}\tau_{3}\Phi$,
which remains well-behaved even in curved or torsion-affected spacetimes.

Beyond its mathematical consistency, the FV transformation preserves
important physical symmetries, allows for a clear separation between
particle and antiparticle degrees of freedom, and admits boundary
conditions and quantization procedures analogous to those in non-relativistic
quantum mechanics. Moreover, it allows one to define conserved observables
and inner products, facilitating quantization and numerical analysis.
These features make it a robust and versatile tool in relativistic
quantum mechanics, particularly for bosonic fields in nontrivial geometries\cite{Bouzenada_2023,Merad_2000,Staudte_1996}.

In this paper, we exploit the strengths of the FV formalism to study
spin-0 particles in the spacetime of a spinning cosmic string endowed
with both angular momentum and axial torsion. The background metric
introduces spacelike disclination (angular deficit) and dislocation
(screw-type torsion), making it a rich testing ground for relativistic
wave equations. By deriving the FV equations in this setting, we obtain
exact analytical solutions for confined particles and analyze their
energy spectra, radial wavefunctions, and Feshbach--Villars densities.
Unlike the KG density, which may turn negative or become non-conserved
in such backgrounds, the FV probability density retains its physical
reliability throughout the entire parameter space.

We further extend the analysis by computing Fisher information \cite{Fisher_1925,Shannon_1948}
and Shannon entropy\cite{Shannon_1948}, leveraging the positive-definiteness
of the FV density to explore the localization and complexity of quantum
states. These information-theoretic measures \cite{Frieden_1989,Boumali_2024,Merad_2000,Bounames_2001,Frieden_1990,Frieden_1992}
would be ill-defined or unreliable in the KG framework, further demonstrating
the FV formalism's superiority for such analyses.

To situate our contribution within the Feshbach-{}-Villars (FV) program
in curved and topologically nontrivial space-times, we build on recent
FV/FVO studies in rotating or non-inertial cosmic-string backgrounds,
cosmic dislocation, Som-{}-Raychaudhuri geometry, Bonnor-{}-Melvin-{}-$\Lambda$,
and Kaluza-{}-Klein settings \cite{Ahmed2025TMP_FermionicBML,Boumali_2024,AhmedBouzenada2024CTP045401,AhmedBouzenada2024EPJP911,AhmedBouzenada2024IJMPA2450032,AhmedBouzenada2024PhysScr065033,Boumali2024RevMexFis050802,Bouzenada2023AOP169479,Bouzenada2023arXiv_2302_13805,Bouzenada2023arXiv_2304_12496,Bouzenada2023NPB116288,Bouzenada2024NPB116682,Bouzenada2024TMP_SomRay}.
Those works typically treat rotation or torsion in isolation, explore
different backgrounds, or assume unbounded radial domains. By contrast,
here we analyze a spinning cosmic string with both disclination (rotation)
and screw-dislocation (torsion)\} simultaneously and impose a finite
cylindrical (hard-wall) boundary, which enforces normalizability and
quantizes the radial wavenumber via Bessel zeros. Within the generalized
FV framework, we employ the positive-definite FV density to maintain
a clean probabilistic interpretation and to compute information-theoretic
indicators (Fisher information and Shannon entropies) directly from
the FV wavefunction. A key technical result is that rotation-{}-torsion
effects enter through an effective angular-momentum index $\nu$ that
mixes $EJ_{t}$ and $kJ_{z}$, thereby lifting the flat-space degeneracy
in a controllable, fine-structure-like manner that collapses in the
flat limit $(\alpha\!\to\!1,\,J_{t},J_{z}\!\to\!0)$. Throughout we
restrict to the positive-energy stationary sector and work in the
weak-field regime (first order in $G$), ensuring analytical transparency
and robustness of the Bessel structure and quantization. The finite-domain
setting further enables parameter-by-parameter comparisons with the
flat-cylinder Klein-{}-Gordon problem, clarifying the distinct roles
of the conical deficit $\alpha$, frame dragging $J_{t}$, and screw
dislocation $J_{z}$. Beyond spectroscopy, we show that increasing
effective confinement enhances Fisher information while reducing position-space
Shannon entropy, linking geometric and torsional features to localization/complexity
measures within the FV representation. Collectively, these points
complement and extend the above literature and, to our knowledge,
provide the first combined rotation-{}-torsion, finite-domain, information-theoretic
analysis of an FVO for scalar particles.

The structure of the paper is as follows. In Section II, we outline
the spacetime geometry of the spinning cosmic string and formulate
the Klein--Gordon equation in this background. Section III introduces
the Feshbach--Villars transformation, derives the corresponding Hamiltonian
operator, and establishes the radial eigenvalue equation. Section
IV provides exact eigensolutions and examines the behavior of probability
densities. Section V focuses on information-theoretic analysis through
Fisher information and Shannon entropy. Finally, we summarize our
findings in the conclusion and highlight prospects for future work
in more generalized geometries or interacting field theories.

\section{Klein Gordon equation in a spinning Cosmic Strings with Spacelike
Disclination and dislocation}

We consider a scalar field $\Phi$ governed by the Klein-Gordon equation
with curvature coupling\cite{Greiner_2000,Sakurai_1967}:
\begin{equation}
\left(\square+m^{2}-\xi\mathcal{R}\right)\Phi(\boldsymbol{x},t)=0,\label{1}
\end{equation}
where $\xi$ is a dimensionless coupling constant and $\mathcal{R}$
is the Ricci scalar.

To evaluate the wave operator $\square\Phi$, we adopt the line element
corresponding to a spinning cosmic string spacetime \cite{Puntigam_1997,Ozdemir_2005,Jusufi_2016}:
\begin{equation}
ds^{2}=-\left(dt+4GJ^{t}d\varphi\right)^{2}+dr^{2}+\alpha^{2}r^{2}d\varphi^{2}+\left(dz+4GJ^{z}d\varphi\right)^{2},\label{2}
\end{equation}
which includes the following geometrical features:
\begin{itemize}
\item An angular deficit parameterized by $\alpha<1$,
\item A screw dislocation effect from torsion encoded in $J^{z}$,
\item Frame dragging from angular momentum represented by $J^{t}$.
\end{itemize}
The spacelike disclination manifests as a conical singularity aligned
along the string’s axis (see Figures 1 and 2 in the referenced work
\cite{Ozdemir_2005}, which depict disclination and dislocation, respectively).
This singularity introduces a deficit angle $\alpha$ in the spatial
plane orthogonal to the string. In simpler terms, the presence of
the cosmic string results in a locally conical geometry---an angular
deficit (or surplus) encircling the string. Accordingly, in classical
general relativity, a straight cosmic string is modeled as a pure
disclination: the spacetime remains flat everywhere except along the
string, where the conical defect resides.

The parameters $J^{t}$ and $J^{z}$, which appear in Equation (\ref{2}),
quantify the intrinsic spin and the extent of spatial dislocation
of the string, respectively. These parameters are reminiscent of torsion
or the Burgers vector in condensed matter physics, which (i) it is
the fundamental topological invariant that characterizes a crystal
dislocation and (ii) characterizes the magnitude and orientation of
lattice distortion essential for describing material defect\foreignlanguage{american}{\textcolor{red}{.
}Physically, $J_{t}$ encodes a time--azimuthal (frame-dragging--like)
coupling associated with the string’s intrinsic rotation, while $J_{z}$
parameterizes an axial screw-dislocation (torsion) along the string.
The conical parameter $0<\alpha<1$ fixes the disclination (deficit
angle). }

Now, we extract the metric components in coordinates $x^{\mu}=(t,r,\varphi,z)$:
\begin{equation}
g_{\mu\nu}=\begin{pmatrix}-1 & 0 & -4GJ^{t} & 0\\
0 & 1 & 0 & 0\\
-4GJ^{t} & 0 & \alpha^{2}r^{2}-16G^{2}(J^{t})^{2}+16G^{2}(J^{z})^{2} & 4GJ^{z}\\
0 & 0 & 4GJ^{z} & 1
\end{pmatrix}.\label{3}
\end{equation}
To first order in $G$, we neglect terms of $\mathcal{O}(G^{2})$,
yielding:
\begin{equation}
g_{\mu\nu}\approx\begin{pmatrix}-1 & 0 & -4GJ^{t} & 0\\
0 & 1 & 0 & 0\\
-4GJ^{t} & 0 & \alpha^{2}r^{2} & 4GJ^{z}\\
0 & 0 & 4GJ^{z} & 1
\end{pmatrix},\quad\sqrt{-g}\approx\alpha r.\label{4}
\end{equation}
This approximation is justified by:
\begin{itemize}
\item The gravitational constant $G$ is extremely small in natural units,
so quadratic terms in $G$ are negligible,
\item The string parameters $J^{t}$ and $J^{z}$ are typically small, corresponding
to Planck-scale effects.
\end{itemize}
Although the Ricci scalar is defined as $\mathcal{R}=g^{\mu\nu}\mathcal{R}_{\mu\nu}$,
in this spacetime geometry the curvature vanishes outside the string
core:
\begin{equation}
\mathcal{R}=0,\quad\text{for }r>0.\label{5}
\end{equation}
The Laplace-Beltrami operator in curved spacetime is:
\begin{equation}
\square\Phi=\frac{1}{\sqrt{-g}}\partial_{\mu}\left(\sqrt{-g}g^{\mu\nu}\partial_{\nu}\Phi\right).\label{6}
\end{equation}
Assuming cylindrical symmetry, we use the separable ansatz:
\begin{equation}
\Phi(t,r,\varphi,z)=e^{-iEt}e^{i\ell\varphi}e^{ikz}R(r).\label{7}
\end{equation}
Substituting into the Klein-Gordon equationwe obtain the radial equation:
\begin{equation}
\frac{1}{r}\frac{d}{dr}\left(r\frac{dR}{dr}\right)+\left[\left(E+\frac{4GJ^{t}\ell}{\alpha^{2}r^{2}}\right)^{2}-\frac{\ell^{2}}{\alpha^{2}r^{2}}-\left(k+\frac{4GJ^{z}\ell}{\alpha^{2}r^{2}}\right)^{2}-m^{2}\right]R=0.\label{8}
\end{equation}
Defining $A=\tfrac{4GJ^{t}\ell}{\alpha^{2}}$ and $B=\tfrac{4GJ^{z}\ell}{\alpha^{2}}$,
this can be rearranged as
\[
R''+\frac{1}{r}R'+\Bigg[\underbrace{(E^{2}-k^{2}-m^{2})}_{\kappa}\;-\;\frac{\underbrace{\ell^{2}/\alpha^{2}-2EA+2kB}_{\mu}}{r^{2}}\;+\;\frac{\underbrace{A^{2}-B^{2}}_{\nu}}{r^{4}}\Bigg]R=0.
\]
The $r^{-4}$ term is $\mathcal{O}(G^{2})$ and renders the equation
non-Bessel: it produces an ODE with irregular singularities at $r=0$
and $r\to\infty$, whose closed-form solutions are of the (double-)confluent
Heun type (HeunD) rather than elementary special functions. Consequently,
exact analytic spectra require handling Heun functions and are generally
intractable for practical boundary conditions. Since the background
itself is kept to leading order in $G$, we consistently drop the
$\nu/r^{4}$ term. 

In that way, retaining only terms up to first order in $G$ and eglecting
the curvature term $\mathcal{R}$ and simplifying further, we arrive
at:
\begin{equation}
\frac{1}{r}\frac{d}{dr}\left(r\frac{dR}{dr}\right)+\left[E^{2}-k^{2}-m^{2}+\frac{8G\ell EJ^{t}}{\alpha^{2}r^{2}}-\frac{\ell^{2}}{\alpha^{2}r^{2}}-\frac{8G\ell kJ^{z}}{\alpha^{2}r^{2}}\right]R=0.\label{9}
\end{equation}
This can be rewritten in standard Bessel form:
\begin{equation}
\frac{d^{2}R}{dr^{2}}+\frac{1}{r}\frac{dR}{dr}+\left[\kappa^{2}-\frac{\nu^{2}}{r^{2}}\right]R=0,\label{10}
\end{equation}
where:
\begin{equation}
\kappa^{2}=E^{2}-k^{2}-m^{2},\quad\nu^{2}=\frac{\ell^{2}}{\alpha^{2}}-\frac{8G\ell}{\alpha^{2}}(EJ^{t}-kJ^{z}).\label{11}
\end{equation}
Or more compactly:
\begin{equation}
\nu=\frac{|\ell|}{\alpha}\left(1-\frac{4G}{|\ell|}(EJ^{t}-kJ^{z})\right).\label{12}
\end{equation}
This is a modified Bessel-type differential equation incorporating
corrections from both torsion ($J^{z}$) and rotation ($J^{t}$) to
first order in $G$\cite{Abramowitz_1964,Andrews_2001,Arfken_2012}.

The general solution is:
\begin{equation}
R(r)=AJ_{\nu}(\kappa r)+BY_{\nu}(\kappa r),\label{13}
\end{equation}
where $J_{\nu}$ and $Y_{\nu}$ are Bessel functions of the first
and second kind, respectively. The parameter $\nu$ contains the combined
effects of the topological defect ($\alpha$), angular momentum ($\ell$),
and first-order gravitational corrections.

To ensure regularity at the origin $r=0$, we discard the divergent
term $Y_{\nu}$, leading to:
\begin{equation}
R(r)=AJ_{\nu}(\kappa r).\label{14}
\end{equation}
The complete scalar field becomes:
\begin{equation}
\Phi(t,r,\varphi,z)=Ae^{-iEt}e^{i\ell\varphi}e^{ikz}J_{\nu}(\kappa r).\label{15}
\end{equation}
To obtain a discrete energy spectrum, we impose a boundary condition
on a finite cylinder of radius $r_{0}$\cite{Fluegge_1994}:
\begin{equation}
\Phi(t,r=r_{0},\varphi,z)=0\quad\Rightarrow\quad J_{\nu}(\kappa r_{0})=0.\label{16}
\end{equation}
Let $j_{\nu,n}$ denote the $n$-th zero of $J_{\nu}$. Then:
\begin{equation}
\kappa_{n}R_{0}=j_{\nu,n}\quad\Rightarrow\quad\kappa_{n}=\frac{j_{\nu,n}}{R_{0}}.\label{17}
\end{equation}
Using the relation $\kappa_{n}^{2}=E_{n}^{2}-k^{2}-m^{2}$, we find:
\begin{equation}
E_{n}=\pm\sqrt{\left(\frac{j_{\nu,n}}{R_{0}}\right)^{2}+k^{2}+m^{2}}.\label{18}
\end{equation}
Thus, the presence of the spinning cosmic string modifies the energy
spectrum via the effective index $\nu$, incorporating both geometric
(conical) and gravitational (torsion and rotation) effects to leading
order in $G$.

At this stage, \foreignlanguage{american}{our metric and wave function
are treated consistently to first order in $G$. Retaining only $\mathcal{O}(G)$
terms yields the Bessel-type radial equation with effective order
$\nu$ quoted above, from which spectra and densities follow analytically.
If one does not truncate the squares in the exact separated equation
\[
\frac{1}{r}\frac{d}{dr}\!\left(r\frac{dR}{dr}\right)+\Bigg[\Big(E+\frac{4GJ_{t}\ell}{\alpha^{2}r^{2}}\Big)^{2}-\frac{\ell^{2}}{\alpha^{2}r^{2}}-\Big(k+\frac{4GJ_{z}\ell}{\alpha^{2}r^{2}}\Big)^{2}-m^{2}\Bigg]R=0,
\]
additional $\mathcal{O}(G^{2})$ terms $\propto r^{-4}$ appear. These
terms move the problem out of the Bessel class, producing a strongly
singular short-distance behavior that requires core regularization
and, in general, alters the quantization condition away from the simple
Bessel-zero condition. Qualitatively, such $G^{2}$ corrections can
(i) generate small energy shifts beyond those captured by the linear
$(EJ_{t}-kJ_{z})$ mixing, and (ii) modify near-axis behavior, potentially
necessitating a refined boundary condition at the string core. }

\section{Dynamics of Feshbach-Villars transformation for scalar particle in
in a spinning Cosmic Strings with Torsion }

We would now want to investigate the quantum dynamics of spin-0 particles
in the space-time caused by a (3+1)-dimensional dislocation, as well
as develop the relevant FV formulation \cite{Feshbach_1958}.

Lets us
\begin{equation}
\Phi=\left(\begin{array}{c}
\phi\\
\chi
\end{array}\right)\label{19}
\end{equation}
In the considered case, the transformation consists in the following
definition of components of the wave function:
\begin{equation}
\begin{gathered}\psi=\phi+\chi,\quad i\left(\partial_{0}+\Upsilon\right)\psi=m(\phi-\chi),\end{gathered}
\label{20}
\end{equation}
with
\begin{equation}
\Upsilon=\frac{1}{2g^{00}\sqrt{-g}}\left\{ \partial_{i},\sqrt{-g}g^{0i}\right\} ,\label{21}
\end{equation}
where $N$ is an arbitrary nonzero real parameter. For the Feshbach-Villars
transformation, it is definite and equal to the particle mass $m$.
This generalization allows us to represent Eq. (6) in the Hamiltonian
form describing both massive and massless particles \cite{Silenko_2008,Silenko_2008_TMP,Silenko_2013}:
\begin{equation}
i\frac{\partial\Phi}{\partial t}=\mathcal{H}\Phi\label{22}
\end{equation}
with
\begin{equation}
\mathcal{H}=\tau_{3}\frac{m^{2}+T}{2m}+i\tau_{2}\frac{-m^{2}+T}{2m}-i\Upsilon\label{23}
\end{equation}
Here
\begin{equation}
T=\frac{1}{g^{00}\sqrt{-g}}\partial_{i}\sqrt{-g}g^{ij}\partial_{j}+\frac{m^{2}}{g^{00}}-\Upsilon^{2}\label{24}
\end{equation}
The incorporation of the spinning cosmic string spacetime metric introduces
significant modifications to the problem. This metric contains off-diagonal
terms and includes additional physical parameters associated with
angular momentum and torsion. As a result, the Klein-Gordon equation,
its solutions, and the corresponding scattering behavior must be reformulated
to incorporate these new geometrical and physical characteristics.

\subsection{Eigensolutions of FV0 in hard wall potential}

We study the quantum dynamics of spin-0 particles propagating in the
curved spacetime generated by a spinning and twisting topological
defect. This background geometry is modeled by a stationary, cylindrically
symmetric metric that generalizes the cosmic string solution by incorporating
both angular momentum and torsion. The line element is given by:
\begin{equation}
ds^{2}=-\left(dt+4GJ^{t}\,d\varphi\right)^{2}+dr^{2}+\alpha^{2}r^{2}\,d\varphi^{2}+\left(dz+4GJ^{z}\,d\varphi\right)^{2}\tag{1}\label{25}
\end{equation}

This spacetime includes two torsional parameters: $J^{t}$, representing
the time--angular momentum coupling, and $J^{z}$, accounting for
axial torsion. The constant $0<\alpha<1$ characterizes the conical
geometry due to the string.

The dynamics of scalar particles in curved spacetime are governed
by the covariant Klein--Gordon equation. We employ the Feshbach--Villars
(FV) formalism to rewrite it in Hamiltonian form using a two-component
wavefunction $\Phi=(\phi,\chi)^{T}$, where the physical scalar field
is $\psi=\phi+\chi$. 

The transformation relating $\phi$, $\chi$, and $\psi$ is:
\begin{equation}
i(\partial_{t}+\Upsilon)\psi=m(\phi-\chi)\label{26}
\end{equation}
\begin{equation}
\Upsilon=-\frac{2GJ^{t}}{\alpha^{2}r^{2}}\partial_{\varphi}\label{27}
\end{equation}
The FV Hamiltonian that governs the time evolution of $\Phi$ is:
\begin{equation}
\mathcal{H}=\tau_{3}\frac{T_{0}+\Delta T}{2m}+i\tau_{2}\frac{T_{0}+\Delta T-2m^{2}}{2m}+i\frac{2GJ^{t}}{\alpha^{2}r^{2}}\partial_{\varphi}\label{28}
\end{equation}
The kinetic and torsional parts of the spatial operator are:
\begin{equation}
T_{0}=-\left[\frac{1}{r}\partial_{r}(r\partial_{r})+\frac{1}{\alpha^{2}r^{2}}\partial_{\varphi}^{2}+\partial_{z}^{2}\right],\quad\Delta T=-\left[\left(\frac{2GJ^{t}}{\alpha^{2}r^{2}}\right)^{2}\partial_{\varphi}^{2}+\frac{8GJ^{z}}{\alpha^{2}r^{2}}\partial_{\varphi}\partial_{z}\right]\label{29}
\end{equation}
The second-order wave equation for the physical scalar field $\psi$
becomes:
\begin{equation}
\left[(\partial_{t}+\Upsilon)^{2}+T\right]\psi=m^{2}\psi\label{30}
\end{equation}
To simplify the analysis, we consider the weak-field regime and linearize
the wave equation by retaining only first-order terms in $G$ (as
the above section). This yields:
\begin{equation}
\left[\partial_{t}^{2}-\frac{4GJ^{t}}{\alpha^{2}r^{2}}\partial_{t}\partial_{\varphi}-\frac{1}{r}\partial_{r}(r\partial_{r})-\frac{1}{\alpha^{2}r^{2}}\partial_{\varphi}^{2}+\partial_{z}^{2}+\frac{8GJ^{z}}{\alpha^{2}r^{2}}\partial_{\varphi}\partial_{z}+m^{2}\right]\psi=0\label{31}
\end{equation}
We assume cylindrical symmetry and use the ansatz:
\begin{equation}
\psi(t,r,\varphi,z)=e^{-iEt}e^{i\ell\varphi}e^{ikz}R(r)\label{32}
\end{equation}
Substituting into the wave equation leads to the radial equation:
\begin{equation}
\left[\frac{d^{2}}{dr^{2}}+\frac{1}{r}\frac{d}{dr}+(E^{2}-m^{2}+k^{2})-\frac{\ell_{\text{eff}}^{2}}{\alpha^{2}r^{2}}\right]R(r)=0\label{33}
\end{equation}
where the effective angular momentum is defined by:
\begin{equation}
\ell_{\text{eff}}^{2}=\ell^{2}-4\alpha^{2}GJ^{t}E\ell+8\alpha^{2}GJ^{z}\ell k\label{34}
\end{equation}
This is a Bessel-type differential equation. The general solution
is:
\begin{equation}
R(r)=AJ_{\nu}(\kappa r)+BY_{\nu}(\kappa r),\quad\nu=\frac{|\ell_{\text{eff}}|}{\alpha},\quad\kappa^{2}=E^{2}-m^{2}+k^{2}\label{35}
\end{equation}
Imposing regularity at the origin requires $B=0$, so the physical
solution becomes:
\begin{equation}
R(r)=AJ_{\nu}(\kappa r)\label{36}
\end{equation}
For a particle confined inside a cylinder of radius $r_{0}$, we require
$R(r_{0})=0$, which leads to:
\begin{equation}
J_{\nu}(\kappa_{n}R_{0})=0\Rightarrow\kappa_{n}=\frac{x_{\nu,n}}{R_{0}},\quad E_{n}=\pm\sqrt{m^{2}-k^{2}+\left(\frac{x_{\nu,n}}{r_{0}}\right)^{2}}\label{37}
\end{equation}

\begin{figure}
\begin{centering}
\includegraphics[scale=0.5]{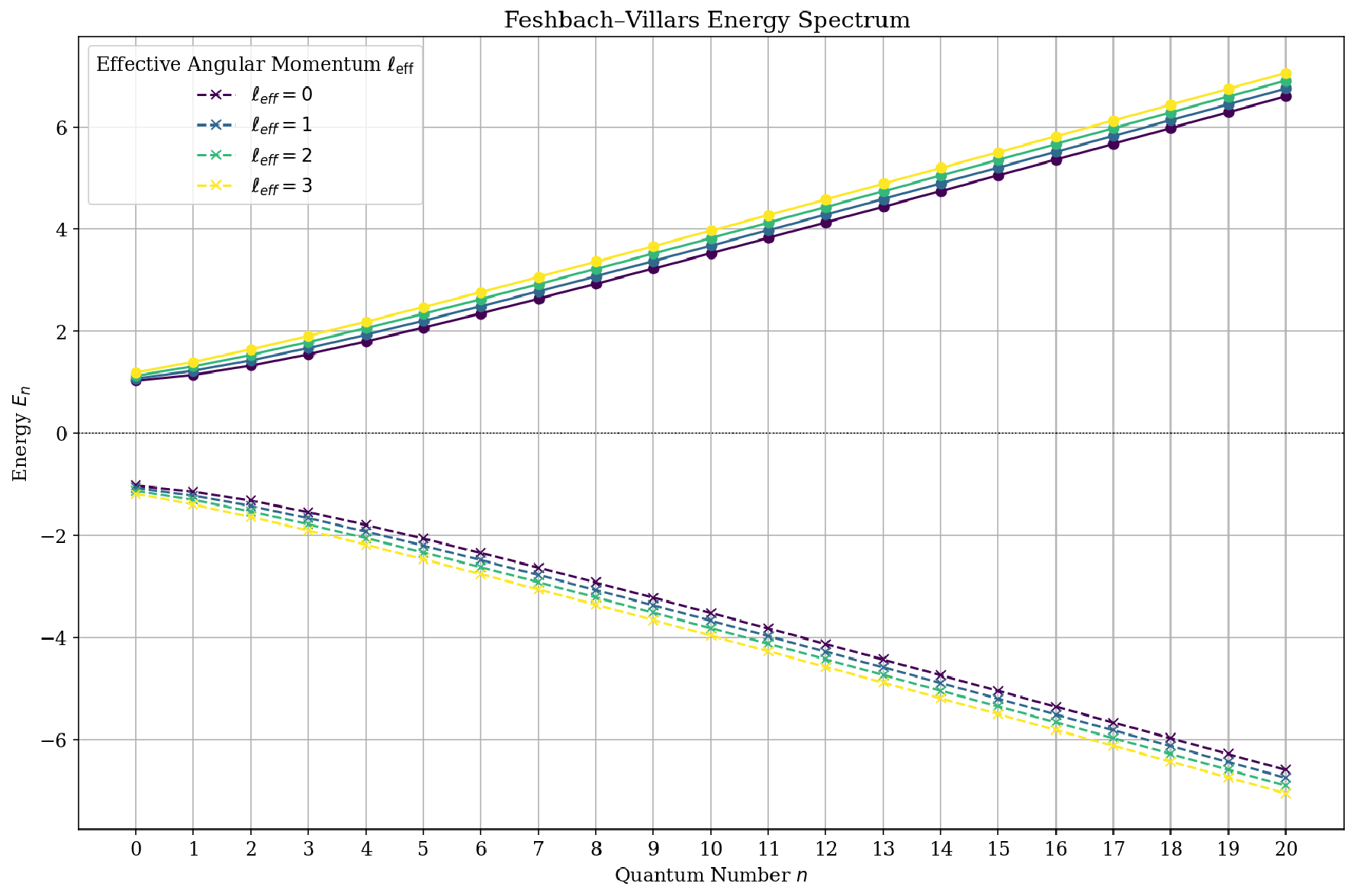}
\par\end{centering}
\caption{\foreignlanguage{american}{Discrete eigen\protect\nobreakdash-energies
$E_{n}$ plotted against the radial quantum number $n$ for several
values of the effective angular momentum $\ell_{\text{eff}}$ in the
spacetime of a spinning cosmic string with torsion.}}\label{fig:1}
\end{figure}

Figures. \ref{fig:1} through. \ref{fig:4} illustrate key physical
insights derived from the Feshbach--Villars (FV) formalism applied
to spin-0 particles in a curved spacetime background induced by a
spinning cosmic string with spacelike disclination and dislocation.
These results emphasize the conceptual and computational advantages
of the FV transformation over the standard Klein--Gordon (KG) framework,
particularly in dealing with relativistic quantum systems in curved
geometries.

To validate our results, we exhibit parameter regimes in which the
spectrum and eigenfunctions reduce to standard Klein--Gordon solutions.\foreignlanguage{american}{
In all cases the radial equation takes the Bessel form
\begin{equation}
R''(r)+\frac{1}{r}R'(r)+\left(\kappa^{2}-\frac{\nu^{2}}{r^{2}}\right)R(r)=0,\label{37-1}
\end{equation}
with solutions $R(r)\propto J_{|\nu|}(\kappa r)$ and hard-wall quantization
$J_{|\nu|}(\kappa R)=0$, i.e.
\begin{equation}
\kappa_{\nu n}=\frac{\alpha_{|\nu|,n}}{R},\label{37-2}
\end{equation}
where $\alpha_{|\nu|,n}$ is the $n$-th zero of $J_{|\nu|}$. The
dispersion reads
\begin{equation}
E_{nk_{z}}^{\;2}=m^{2}+\kappa_{\nu n}^{\,2}+k_{z}^{2}.\label{37-3}
\end{equation}
Now,}
\selectlanguage{american}%
\begin{itemize}
\item For a Pure rotation $(\ensuremath{J_{z}=0})$. With screw dislocation
switched off, the effective index reduces to a rotation-only form
$\nu\to\nu_{{\rm rot}}(\alpha,J_{t})$. The spectrum follows from
\begin{equation}
\kappa_{\nu_{{\rm rot}}n}=\alpha_{|\nu|,n}/R\label{37-4}
\end{equation}
This reproduces the known rotating-conical background and its defect-induced
splitting of the $\pm m$ multiplets.
\item For a Pure screw dislocation $(\ensuremath{J_{t}=0}$). Setting the
rotation to zero yields $\nu\to\nu_{{\rm tors}}(\alpha,J_{z})$. The
eigenfunctions remain Bessel functions with the same boundary condition,
recovering the standard static torsional (cosmic dislocation) case.
\item for the flat-spacetime limit $(\ensuremath{\alpha\to1}$, $J_{t}\to0$,
$\ensuremath{J_{z}\to0}).$ In this limit $\nu\to\ell$ and the spectrum
collapses to the textbook cylindrical Klein--Gordon result,
\begin{equation}
\kappa_{n}=\frac{\alpha_{|\nu|,n}}{R},\qquad E_{nk_{z}}^{\;2}=m^{2}+\left(\frac{\alpha_{|\nu|,n}}{R}\right)^{\!2}+k_{z}^{2},\label{37-5}
\end{equation}
\end{itemize}
Finally, we have imposed a hard-wall boundary at radius $R_{0}$ (Dirichlet),
$R(R_{0})=0$, which yields the quantization condition $J_{\nu}(\kappa R_{0})=0$
and the spectrum $E_{n}^{2}=m^{2}+k^{2}+(j_{\nu,n}/R_{0})^{2}$.

The underlying radial equation,
\begin{equation}
R''+\frac{1}{r}R'+\big(\kappa^{2}-\nu^{2}/r^{2}\big)R=0\label{37-6}
\end{equation}
 with $\nu$ defined in Eq. (\ref{35}), has Bessel form; consequently,
changing the boundary condition at $r=R_{0}$ modifies only the quantization
condition and not the bulk differential operator.
\begin{itemize}
\item For Dirichlet (hard wall), $R(R_{0})=0$ implies $J_{\nu}(\kappa R_{0})=0$
.
\item For Neumann, $R'(R_{0})=0$ implies $J'_{\nu}(\kappa R_{0})=0$; the
corresponding zeros lie below the Dirichlet zeros for fixed $\nu$,
leading to slightly lower eigenvalues (a mild compression of the spectrum). 
\item The Robin condition, $R'(R_{0})+\lambda R(R_{0})=0$, gives $J'_{\nu}(\kappa R_{0})+\lambda R_{0}\,J_{\nu}(\kappa R_{0})=0$
and interpolates continuously between the Neumann ($\lambda\to0$)
and Dirichlet ($\lambda R_{0}\to\infty$) limits. 
\end{itemize}
In all cases, the geometric/torsional physics and the associated degeneracy
lifting are governed by the index $\nu(\alpha,J_{t},J_{z};E,k)$;
boundary conditions merely shift the root sequence without altering
this dependence. Hence the level splittings reported in Fig. \ref{fig:1}
are robust with respect to boundary choice, 

\selectlanguage{english}%
Now we are ready to discuss our results. 

Figure. \ref{fig:1} displays the energy spectrum $E_{n}$ of the
Feshbach--Villars oscillator as a function of the radial quantum
number $n$ for several values of the effective angular momentum $\ell_{\text{eff}}$.
The eigenvalues are determined by imposing boundary conditions on
the modified Bessel equation that governs the radial behavior of the
FV wavefunction. The influence of the geometric and torsional parameters
is embedded in $\ell_{\text{eff}}$, which accounts for corrections
from both frame dragging and torsion induced by the cosmic string
spacetime. The clear and systematic shift of the energy levels with
increasing angular momentum confirms the physical relevance of these
geometric contributions. Notably, the spectrum is symmetric with respect
to positive and negative energies, a hallmark of relativistic theories
capable of simultaneously describing particles and antiparticles.
The FV formalism handles this naturally through its two-component
wavefunction, in contrast to the KG equation where interpretation
of negative energy solutions remains problematic.

The FV components $\phi$ and $\chi$ in terms of $\psi$ are:
\begin{equation}
\phi=\frac{1}{2}\left(1+\frac{E}{m}+i\frac{2GJ^{t}\ell}{m\alpha^{2}r^{2}}\right)\psi,\quad\chi=\frac{1}{2}\left(1-\frac{E}{m}-i\frac{2GJ^{t}\ell}{m\alpha^{2}r^{2}}\right)\psi\label{38}
\end{equation}
The Feshbach--Villars observable (FVO) density is computed as:
\begin{equation}
\rho_{\text{FVO}}=\Phi^{\dagger}\tau_{3}\Phi=|\phi|^{2}-|\chi|^{2}=\frac{A^{2}|E|}{m}\left|J_{\nu}(\kappa_{n}r)\right|^{2}\label{39}
\end{equation}
The normalization constant $A$ is obtained as follows. Immediately
after Eq. (\ref{39}) we include
\begin{equation}
A^{-2}=\frac{|E|}{m}\int_{0}^{R_{0}}r\,\big|J_{\nu}(\kappa_{n}r)\big|^{2}\,dr=\frac{|E|}{m}\,\frac{R_{0}^{2}}{2}\,\big[J_{\nu+1}(j_{\nu,n})\big]^{2},\label{39-1}
\end{equation}
using the standard identity at Bessel zeros $j_{\nu,n}$,
\begin{equation}
\int_{0}^{R_{0}}r\,J_{\nu}^{2}(\kappa_{n}r)\,dr=\tfrac{R_{0}^{2}}{2}\,[J_{\nu+1}(j_{\nu,n})]^{2}\label{39-2}
\end{equation}

\begin{figure}
\begin{centering}
\includegraphics[scale=0.4]{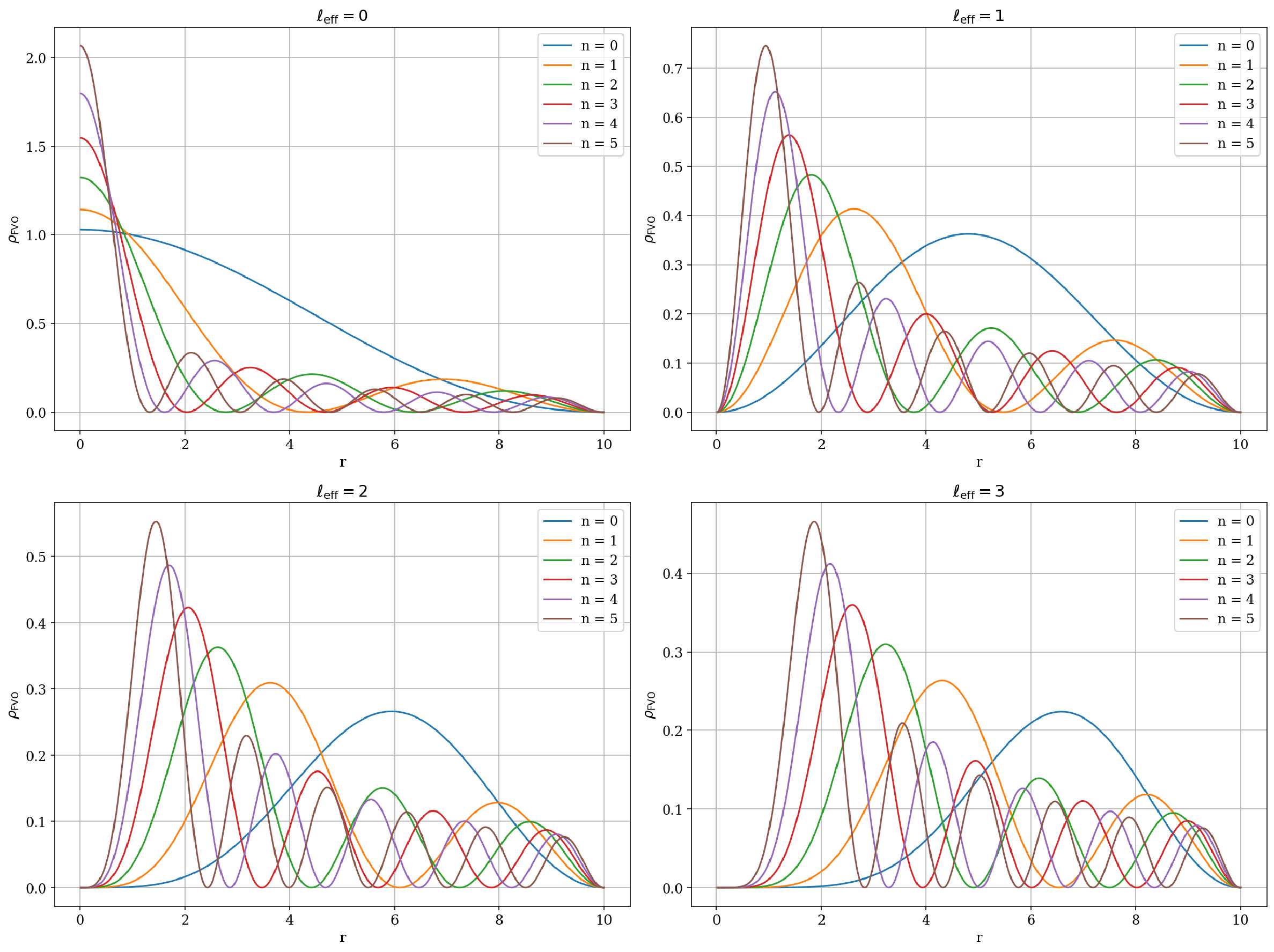}
\par\end{centering}
\caption{\foreignlanguage{american}{Positive\protect\nobreakdash-definite
Feshbach--Villars probability density $\rho_{\text{FVO}}(r)$ for
low\protect\nobreakdash-lying states ($n=0\!-\!5$) at different
$\ell_{\text{eff}}$, showing how torsion and rotation shift the spatial
localization of the wave\protect\nobreakdash-function.}}\label{fig:2}
\end{figure}

\selectlanguage{american}%
We are now in a position to examine the radial probability-density
curves.

Figure. \ref{fig:2} shows the radial probability density of the Feshbach--Villars
(FVO) for spin-0 particles within the spacetime geometry induced by
a spinning cosmic string characterized by spacelike disclination and
dislocation.\foreignlanguage{english}{. The plots clearly demonstrate
that the density remains positive-definite, even for excited states,
which is not guaranteed in the KG framework due to its dependence
on second-order time derivatives. In the KG case, the probability
density includes the time derivative of the wavefunction, allowing
for negative values and complicating the physical interpretation.
In contrast, the FV approach yields a conserved and well-behaved density
that facilitates consistent probabilistic interpretation, making it
particularly suitable for relativistic quantum systems in nontrivial
geometries.}

Importantly, all probability densities presented remain strictly positive.
This highlights a crucial advantage of the FV formalism over the Klein--Gordon
(KG) equation, as the latter can yield negative densities due to its
second-order time derivatives. Specifically, in the KG framework,
the density depends explicitly on the wavefunction and its time derivative
, allowing arbitrary instantaneous values and thus potentially resulting
in negative densities. Therefore, the positive-definite nature of
the FV probability density ensures a well-defined probabilistic interpretation,
making the FV oscillator particularly effective for investigating
quantum phenomena in curved spacetimes and gravitational fields associated
with cosmic strings and other topological defects.

\selectlanguage{english}%
All the presented figures demonstrate that the probability density
in the Feshbach--Villars (FV) formalism remains strictly positive,
unlike in the Klein--Gordon (KG) equation, where the density may
become negative. This discrepancy stems from the fact that the KG
equation is second-order in time, leading to a probability density
that involves a time derivative of the wavefunction. Since both $\psi$
and $\partial\psi/\partial t$ can take arbitrary values at a given
instant, the resulting density $\rho$ is not guaranteed to be positive-definite
and can assume negative values.

In contrast, the FV formalism reformulates the dynamics into a first-order
time evolution equation, resulting in a two-component wavefunction
and a probability density that is positive-definite and well-suited
for probabilistic interpretation. The FV probability density is explicitly
given by:
\begin{equation}
\rho_{\mathrm{FVO}}=\Phi^{\dagger}\tau_{3}\Phi\label{eq:40}
\end{equation}
as shown in Equation (\ref{39}), where $\Phi$ contains the particle
and antiparticle components $\phi$ and $\chi$, respectively. The
conserved total charge in this formulation is then:
\begin{equation}
Q=\int\rho_{\mathrm{FVO}}\,d^{3}x=\pm1\label{eq:41}
\end{equation}
This charge conservation reflects the contributions from both particles
and antiparticles, providing a robust framework to interpret negative-energy
solutions that arise naturally in relativistic quantum theory.

\section{Fisher and Entropy information for FV0 particles }

Furthermore, a significant advantage of the FV formalism is that information-theoretic
quantities such as the Fisher information and Shannon entropy can
be directly calculated from $\rho_{\mathrm{FVO}}$ due to its regularity
and positivity. These quantities are defined as:
\begin{itemize}
\item Fisher Information:
\begin{equation}
I_{r}=\int\frac{1}{\rho(r)}\left(\frac{d\rho(r)}{dr}\right)^{2}dv\label{eq:42}
\end{equation}
\item Shannon Entropy:
\begin{equation}
S_{r}=-\int\rho(r)\ln\rho(r)\,dv\label{eq:43}
\end{equation}
\end{itemize}
These measures provide insights into the spread, localization, and
information content of quantum states. Such direct computations are
not feasible in the KG framework due to the indefinite sign of the
density and its dependence on time derivatives, which compromise the
interpretation of $\rho$ as a genuine probability distribution.

In summary, the Feshbach--Villars formalism not only resolves the
issue of negative probabilities found in KG theory but also enables
meaningful and physically consistent calculations of statistical and
informational properties of relativistic quantum systems.

\begin{figure}
\begin{centering}
\includegraphics[scale=0.4]{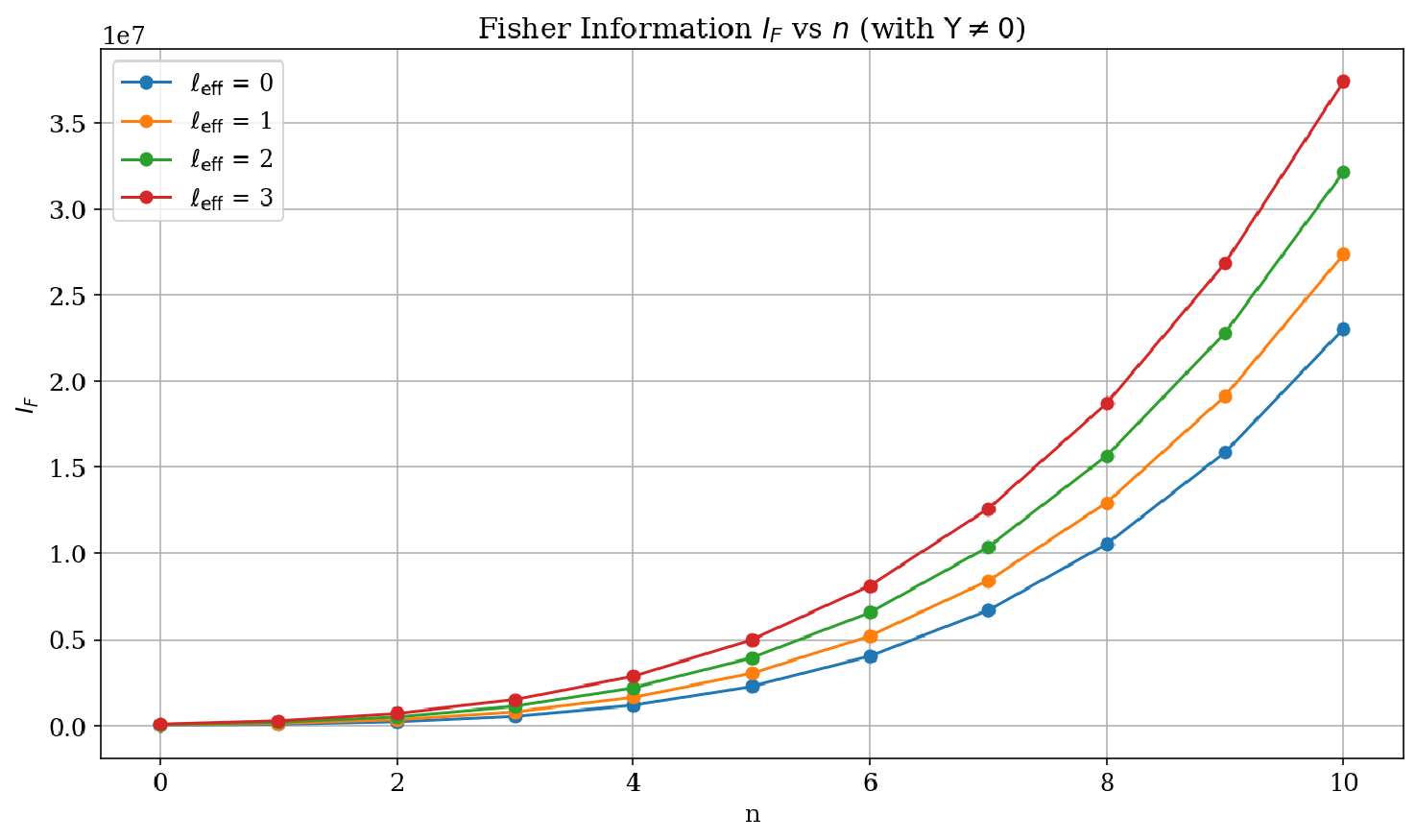}
\par\end{centering}
\caption{\foreignlanguage{american}{Variation of the Fisher information $I_{F}$
with radial quantum number $n$ for several $\ell_{\text{eff}}$;
higher $I_{F}$ at large $n$ indicates increasing localization of
the scalar particle in the defect space--time.}}\label{fig:3}
\end{figure}

\begin{figure}
\begin{centering}
\includegraphics[scale=0.4]{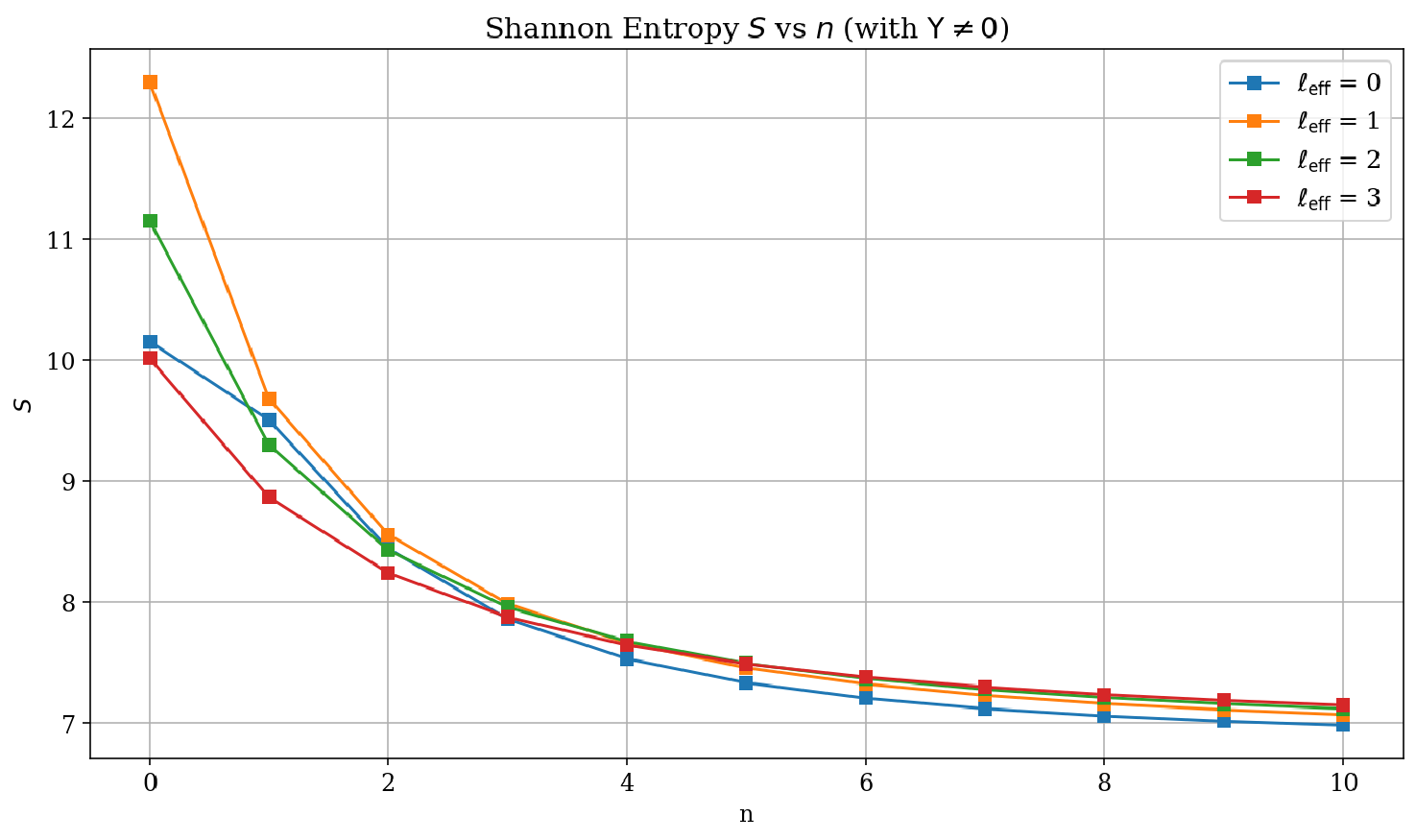}
\par\end{centering}
\caption{\foreignlanguage{american}{Shannon entropy $S$ as a function of
$n$ for different $\ell_{\text{eff}}$; the declining trend reflects
the reduction in configurational complexity as the states become more
localized.}}\label{fig:4}
\end{figure}

Figure. \ref{fig:3}shows the Fisher information $I_{r}$ as a function
of the quantum number $n$, reflecting the localization characteristics
of the FV wavefunctions. As $n$ increases, the wavefunctions exhibit
sharper spatial features, and $I_{r}$ rises accordingly, especially
for higher $\ell_{\text{eff}}$. This monotonic behavior indicates
an increasing spatial resolution of quantum states with higher energy,
a feature that is physically meaningful only when the underlying probability
density is well-defined and positive. Such an analysis would be ill-posed
in the KG framework, where the presence of indefinite densities makes
the computation of information-theoretic quantities unreliable.

Figure. \ref{fig:4} complements this by showing the Shannon entropy
$S_{r}$ as a function of $n$, which decreases with increasing quantum
number, indicating a trend toward more localized states. The entropy
curves exhibit systematic behavior across different values of $\ell_{\text{eff}}$,
again underscoring the robust statistical interpretability of the
FV density. This reinforces the utility of the FV formalism in evaluating
both the spatial complexity and informational content of quantum states
in relativistic systems, a task fundamentally compromised in the KG
theory due to its indefinite norm.

Together, these figures underscore the practical and conceptual benefits
of the FV transformation in modeling spin-0 particles under curved
spacetime backgrounds with topological defects. By reformulating the
second-order KG equation into a first-order Hamiltonian form with
a two-component wavefunction, the FV formalism enables positive-definite
densities, clear separation of particle and antiparticle states, and
the possibility of computing meaningful statistical and information-theoretic
measures. This makes it a powerful and consistent tool for probing
quantum effects in gravitational and topologically nontrivial settings.

\subsection{Information-theoretic interpretation}

Let $\rho(\mathbf{r})$ and $\gamma(\mathbf{p})$ denote the position-
and momentum-space probability densities ($\hbar=1$). Their differential
(Shannon) entropies are
\begin{equation}
S_{r}=-\!\int\rho\ln\rho\,d^{D}r,\qquad S_{p}=-\!\int\gamma\ln\gamma\,d^{D}p,\label{eq:44}
\end{equation}
and the corresponding Fisher informations, which quantify the local
gradient content (sharpness) of the distributions, are
\begin{equation}
I_{r}[\rho]=\int_{\mathbb{R}^{D}}\frac{|\nabla\rho(\mathbf{r})|^{2}}{\rho(\mathbf{r})}\,d^{D}r=4\!\int_{\mathbb{R}^{D}}\!|\nabla\sqrt{\rho}|^{2}\,d^{D}r,\qquad I_{p}=\int_{\mathbb{R}^{D}}\frac{|\nabla\gamma|^{2}}{\gamma}\,d^{D}p.\label{eq:45}
\end{equation}
Introducing the entropy power
\begin{equation}
N_{r}=\frac{1}{2\pi e}\exp\!\Big(\frac{2}{D}S_{r}\Big),\label{eq:46}
\end{equation}
the Stam/Cramér--Rao chain yields \cite{Stam1959,Cramer1959}
\begin{equation}
N_{r}\,I_{r}\;\ge\;D,\label{eq:47}
\end{equation}
with equality for Gaussians (and an analogous momentum-space statement
$N_{p}\,I_{p}\ge D$). Thus, a reduction of $S_{r}$ (stronger real-space
localization) necessarily increases $I_{r}$. In parallel, the entropic
uncertainty relation (EUR or BBM relation \cite{BBM1975}) and the
Fisher uncertainty relation,
\begin{equation}
S_{r}+S_{p}\;\ge\;D\,(1+\ln\pi),\qquad I_{r}\,I_{p}\;\ge\;4D^{2},\label{eq:48}
\end{equation}
enforce the position--momentum trade-off: localization in configuration
space (lower $S_{r}$, higher $I_{r}$) must be compensated by delocalization
or increased structure in momentum space (higher $S_{p}$, constrained
$I_{p}$). Physically, $I_{r}$ coincides (up to constants) with the
Weizsäcker inhomogeneity contribution to the kinetic energy; increased
nodal structure or steeper spatial gradients raise this minimal kinetic-energy
content. We note here that the Weizsäcker inhomogeneity contribution
is a fundamental component in density functional theory (DFT), introduced
by Carl Friedrich von Weizsäcker in 1935 \cite{Weizsaecker1935} to
refine the kinetic energy functional for systems with spatially varying
electron density $\ensuremath{\rho(\mathbf{r})}.$ It accounts for
the kinetic energy arising from density gradients, which is neglected
in the uniform electron gas approximation of the Thomas-Fermi model. 

The Weizsäcker term is given by:
\begin{equation}
T_{W}[\rho]=\frac{\hbar^{2}}{8m}\int\frac{|\nabla\rho(\mathbf{r})|^{2}}{\rho(\mathbf{r})}\,d^{D}\mathbf{r},\label{eq:49}
\end{equation}
This expression is directly proportional to the Fisher information
\begin{equation}
I_{r}=\int\frac{|\nabla\rho|^{2}}{\rho}\,d^{D}\mathbf{r}\label{49}
\end{equation}
, capturing the local inhomogeneity or \textquotedbl sharpness\textquotedbl{}
of the density distribution. The term is critical in regions of rapid
density variation, such as near atomic cores or in confined quantum
systems, and enhances the accuracy of kinetic energy functionals in
DFT. It also connects to information-theoretic measures, supporting
the interplay between position-space localization and momentum-space
delocalization as governed by quantum uncertainty principles.

The monotonic increase of $I_{r}(n)$ with the radial quantum number
$n$ (more nodes $\Rightarrow$ steeper $\nabla\rho$) and the simultaneous
decrease of $S_{r}(n)$ observed in Figs. 3--4 are therefore mutually
consistent consequences of the bounds above: the Stam inequality explains
why sharper spatial structure forces $I_{r}$ upward, while the EUR
guarantees a compensating broadening in momentum space (larger $S_{p}$).

Figure 4 shows a decreasing $S_{r}$ with increasing $n$, indicating
enhanced localization of the bound states; this behavior persists
in related geometries \cite{Moreira2025GodelEntropy,Yang2022CosmicStringQFI,Arvizu2024ConicalEntropy,Lima2023QIE_AB}:
\begin{itemize}
\item Cosmic string without spin/torsion. In the spinless conical background
(deficit parameter $\alpha<1$), the azimuthal sector is effectively
compressed, tightening radial confinement. As \$\textbackslash alpha\$
decreases, $S_{r}$ is reduced while $I_{r}$ grows, reproducing the
qualitative tendency reported for the pure cosmic-string case .
\item Gödel-type metrics (frame dragging). The rotation parameter acts as
a magnetic-like coupling that generates Landau-type radial confinement.
Increasing rotation strengthens real-space localization, thereby lowering
$S_{r}$ and raising $I_{r}$. Although degeneracy patterns differ
from the conical case, the monotonic entropy trend aligns with the
intuition of rotation-induced confinement .
\end{itemize}
Finally,\foreignlanguage{american}{ the observed trends $I_{r}(n)$
increasing and $S_{r}(n)$ decreasing---quantify sharpening real-space
structure (more nodes $\Rightarrow$ steeper $\nabla\sqrt{\rho}$)
and reduced configurational complexity. The Stam/Cramér--Rao chain
and entropic uncertainty (BBM) guarantee that reduced $S_{r}$ must
be compensated by increased momentum-space complexity $S_{p}$ and/or
$I_{p}$, consistent with our spectra in a confining conical/torsional
geometry. Thus, Fisher/entropy serve as compact proxies for “localization
vs. complexity” in this background and are natural diagnostics for
relativistic quantum dynamics with defects. }

\section{Conclusion }

\selectlanguage{american}%
We have developed a Feshbach--Villars (FV) treatment of scalar quantum
dynamics in the spacetime of a spinning cosmic string with simultaneous
disclination (deficit $\alpha$) and screw-dislocation ($J_{z}$),
including frame dragging ($J_{t}$). In the consistent $\mathcal{O}(G)$
approximation, the separated radial problem assumes Bessel form with
an effective angular-momentum index $\nu(\alpha,J_{t},J_{z};E,k)$
that encodes a rotation--torsion mixing proportional to $(EJ_{t}-kJ_{z})$.
Imposing a cylindrical hard-wall at $R_{0}$ produces the simple quantization
$E_{n}=m^{2}+k^{2}+(j_{\nu,n}/R_{0})^{2}$, from which we constructed
normalized eigenfunctions and strictly positive FV densities. The
FV formalism thereby circumvents the sign-indefinite Klein--Gordon
density and permits direct computation of information-theoretic quantities:
the Fisher information increases and the position-space Shannon entropy
decreases with stronger effective confinement and with radial quantum
number, consistent with Stam/Cramér--Rao and entropic-uncertainty
relations. Limiting cases (pure rotation, pure torsion, and the flat
limit $\alpha\to1$,$J_{t},J_{z}\to0$) are recovered smoothly, validating
the framework.

Two technical boundaries of our analysis are worth highlighting. First,
retaining the full squared terms in the exact radial equation introduces
$\mathcal{O}(G^{2})$, $r^{-4}$ contributions that move the problem
outside the Bessel class into (double)-confluent Heun territory and
require core regularization; our weak-field results therefore delineate
the analytically tractable regime. Second, while Dirichlet confinement
was emphasized, Neumann/Robin or finite-step boundaries shift the
root conditions without altering the underlying $\nu$-driven geometry/torsion
dependence.

The FV approach thus offers a precise, physically transparent baseline
for scalar spectroscopy and information measures in rotating/torsional
string spacetimes. Natural extensions include: (i) controlled inclusion
of $\mathcal{O}(G^{2})$ effects and core models to quantify Heun-level
corrections; (ii) external electromagnetic fields and finite-step/soft
confinements; (iii) time-dependent or non-stationary backgrounds;
and (iv) momentum-space entropy/Fisher analyses and scattering observables
to connect with transport in defected media.

\selectlanguage{english}%
Recent progress has moved the Dirac oscillator from a purely theoretical
construct to an experimentally accessible system: a one-dimensional
implementation was realized using a chain of coupled microwave resonators
that emulates the tight-binding form of the Dirac oscillator. Complementary
demonstrations of Dirac dynamics have been achieved with trapped ions,
which faithfully simulate relativistic wave-packet motion, and in
condensed-matter “Dirac materials” such as graphene, whose quasiparticles
obey effective massless Dirac equations \cite{moshinsky1989,Lamata2011,Blatt2012,FrancoVillafane2013DiracOscillator,Gerritsma2010DiracEquationIons}.

\bibliographystyle{ChemEurJ}
\bibliography{referencesara}

\providecommand{\url}[1]{\texttt{#1}}
\providecommand{\urlprefix}{}
\providecommand{\foreignlanguage}[2]{#2}
\providecommand{\Capitalize}[1]{\uppercase{#1}}
\providecommand{\capitalize}[1]{\expandafter\Capitalize#1}
\providecommand{\bibliographycite}[1]{\cite{#1}}
\providecommand{\bbland}{and}
\providecommand{\bblchap}{chap.}
\providecommand{\bblchapter}{chapter}
\providecommand{\bbletal}{et~al.}
\providecommand{\bbleditors}{editors}
\providecommand{\bbleds}{eds.}
\providecommand{\bbleditor}{editor}
\providecommand{\bbled}{ed.}
\providecommand{\bbledition}{edition}
\providecommand{\bbledn}{ed.}
\providecommand{\bbleidp}{page}
\providecommand{\bbleidpp}{pages}
\providecommand{\bblerratum}{erratum}
\providecommand{\bblin}{in}
\providecommand{\bblmthesis}{Master's thesis}
\providecommand{\bblno}{no.}
\providecommand{\bblnumber}{number}
\providecommand{\bblof}{of}
\providecommand{\bblpage}{page}
\providecommand{\bblpages}{pages}
\providecommand{\bblp}{p}
\providecommand{\bblphdthesis}{Ph.D. thesis}
\providecommand{\bblpp}{pp}
\providecommand{\bbltechrep}{Tech. Rep.}
\providecommand{\bbltechreport}{Technical Report}
\providecommand{\bblvolume}{volume}
\providecommand{\bblvol}{Vol.}
\providecommand{\bbljan}{January}
\providecommand{\bblfeb}{February}
\providecommand{\bblmar}{March}
\providecommand{\bblapr}{April}
\providecommand{\bblmay}{May}
\providecommand{\bbljun}{June}
\providecommand{\bbljul}{July}
\providecommand{\bblaug}{August}
\providecommand{\bblsep}{September}
\providecommand{\bbloct}{October}
\providecommand{\bblnov}{November}
\providecommand{\bbldec}{December}
\providecommand{\bblfirst}{First}
\providecommand{\bblfirsto}{1st}
\providecommand{\bblsecond}{Second}
\providecommand{\bblsecondo}{2nd}
\providecommand{\bblthird}{Third}
\providecommand{\bblthirdo}{3rd}
\providecommand{\bblfourth}{Fourth}
\providecommand{\bblfourtho}{4th}
\providecommand{\bblfifth}{Fifth}
\providecommand{\bblfiftho}{5th}
\providecommand{\bblst}{st}
\providecommand{\bblnd}{nd}
\providecommand{\bblrd}{rd}
\providecommand{\bblth}{th}
\begin{thebibliography}{10}

\bibitem{Vilenkin_1981}
A.~Vilenkin, \emph{Phys. Rev. D} \textbf{1981}, \emph{23}, 852.

\bibitem{Vilenkin_1994}
A.~Vilenkin, E.~P.~S. Shellard, \emph{Cosmic Strings and Other Topological
  Defects}, Cambridge University Press, Cambridge, \textbf{1994}.

\bibitem{Katanaev_1999}
M.~O. Katanaev, I.~V. Volovich, \emph{Annals of Physics} \textbf{1999},
  \emph{271}, 203--232.

\bibitem{Marques_2002}
G.~D.~A. Marques, V.~B. Bezerra, \emph{Phys. Rev. D} \textbf{2002}, \emph{66},
  105011.

\bibitem{Marques_2005}
G.~de~A.~Marques, V.~B. Bezerra, C.~Furtado, F.~Moraes, \emph{Int. J. Mod.
  Phys. A} \textbf{2005}, \emph{20}, 6051.

\bibitem{Boumali_2014}
A.~Boumali, N.~Messai, \emph{Can. J. Phys.} \textbf{2014}, \emph{92}, 1460.

\bibitem{Chen_2020}
H.~Chen, Z.~W. Long, Q.~K. Ran, Y.~Yang, C.~Y. Long, \emph{EPL} \textbf{2020},
  \emph{132}, 50006.

\bibitem{Messai_2015}
N.~Messai, A.~Boumali, \emph{Eur. Phys. J. Plus} \textbf{2015}, \emph{130},
  140.

\bibitem{Boumali_2017}
A.~Boumali, N.~Messai, \emph{Can. J. Phys.} \textbf{2017}, \emph{95}, 999.

\bibitem{Bakke_2009}
K.~Bakke, C.~Furtado, \emph{Eur. Phys. J. C} \textbf{2009}, \emph{69}, 531.

\bibitem{Bakke_2010}
K.~Bakke, C.~Furtado, \emph{Int. J. Mod. Phys. D} \textbf{2010}, \emph{19}, 85.

\bibitem{Bakke_2013}
K.~Bakke, \emph{Gen. Relativ. Gravit.} \textbf{2013}, \emph{45}, 1847.

\bibitem{Garcia_2017}
G.~Q. Garcia, J.~R. de~S.~Oliveira, K.~Bakke, C.~Furtado, \emph{Eur. Phys. J.
  Plus} \textbf{2017}, \emph{132}, 123.

\bibitem{Mota_2017}
H.~F. Mota, K.~Bakke, \emph{Gen. Relativ. Gravit.} \textbf{2017}, \emph{49},
  104.

\bibitem{Castro_2016}
L.~B. Castro, \emph{Eur. Phys. J. C} \textbf{2016}, \emph{76}, 61.

\bibitem{Jaszek_2001}
R.~Jaszek, \emph{J. Mater. Sci. Mater. Electron.} \textbf{2001}, \emph{12}, 1.

\bibitem{VanHove_1954}
L.~V. Hove, \emph{Phys. Rev.} \textbf{1954}, \emph{95}, 249.

\bibitem{Messai_2025}
N.~Messai, A.~Boumali, \emph{European Physical Journal Plus} \textbf{2025},
  \emph{140}, 560.

\bibitem{Dirac_1981}
P.~A.~M. Dirac, \emph{The Principles of Quantum Mechanics} \emph{\bblfourtho{}
  \bbledn{}}, Oxford University Press, \textbf{1988}.

\bibitem{Greiner_2000}
W.~Greiner, \emph{Relativistic Quantum Mechanics: Wave Equations}
  \emph{\bblthirdo{} \bbledn{}}, Springer, \textbf{2000}.

\bibitem{Sakurai_1967}
J.~J. Sakurai, \emph{Advanced Quantum Mechanics}, Addison-Wesley,
  \textbf{1967}.

\bibitem{Feshbach_1958}
H.~Feshbach, F.~M.~H. Villars, \emph{Reviews of Modern Physics} \textbf{1958},
  \emph{30}, 24.

\bibitem{Bouzenada_2023}
A.~Bouzenada, A.~Boumali, \emph{Annals of Physics} \textbf{2023}, \emph{452},
  169302.

\bibitem{Merad_2000}
M.~Merad, L.~Chetouani, A.~Bounames, \emph{Physics Letters A} \textbf{2000},
  \emph{267}, 225.

\bibitem{Staudte_1996}
D.~S. Staudte, \emph{Journal of Physics A} \textbf{1996}, \emph{29}, 169.

\bibitem{Fisher_1925}
R.~A. Fisher, \emph{Proceedings of the Cambridge Philosophical Society}
  \textbf{1925}, \emph{22}, 700.

\bibitem{Shannon_1948}
C.~E. Shannon, \emph{Bell System Technical Journal} \textbf{1948}, \emph{27},
  379--423.

\bibitem{Frieden_1989}
B.~R. Frieden, \emph{Optics Letters} \textbf{1989}, \emph{14}, 199.

\bibitem{Boumali_2024}
A.~Boumali, A.~Hamla, Y.~Chargui, \emph{International Journal of Theoretical
  Physics} \textbf{2024}, \emph{63}.

\bibitem{Bounames_2001}
A.~Bounames, L.~Chetouani, \emph{Physics Letters A} \textbf{2001}, \emph{279},
  139.

\bibitem{Frieden_1990}
B.~R. Frieden, \emph{Physical Review A} \textbf{1990}, \emph{41}, 4265.

\bibitem{Frieden_1992}
B.~R. Frieden, \emph{Physica A} \textbf{1992}, \emph{180}, 359.

\bibitem{Ahmed2025TMP_FermionicBML}
F.~Ahmed, N.~Candemir, A.~Bouzenada, \emph{Theoretical and Mathematical
  Physics} \textbf{2025}, \emph{222}, 170.

\bibitem{AhmedBouzenada2024CTP045401}
F.~Ahmed, A.~Bouzenada, \emph{Communications in Theoretical Physics}
  \textbf{2024}, \emph{76}, 045401.

\bibitem{AhmedBouzenada2024EPJP911}
F.~Ahmed, A.~Bouzenada, \emph{The European Physical Journal Plus}
  \textbf{2024}, \emph{139}, 911.

\bibitem{AhmedBouzenada2024IJMPA2450032}
F.~Ahmed, A.~Bouzenada, \emph{International Journal of Modern Physics A}
  \textbf{2024}, \emph{39}, 2450032.

\bibitem{AhmedBouzenada2024PhysScr065033}
F.~Ahmed, A.~Bouzenada, \emph{Physica Scripta} \textbf{2024}, \emph{99},
  065033.

\bibitem{Boumali2024RevMexFis050802}
A.~Boumali, A.~Bouzenada, N.~Messai, O.~Mustafa, \emph{Revista Mexicana de
  F{\'\i}sica} \textbf{2024}, \emph{70}, 050802.

\bibitem{Bouzenada2023AOP169479}
A.~Bouzenada, A.~Boumali, E.~O. Silva, \emph{Annals of Physics} \textbf{2023},
  \emph{458}, 169479.

\bibitem{Bouzenada2023arXiv_2302_13805}
A.~Bouzenada, A.~Boumali, M.~Al-Raeei, \emph{arXiv preprint} \textbf{2023}.

\bibitem{Bouzenada2023arXiv_2304_12496}
A.~Bouzenada, A.~Boumali, \emph{arXiv preprint} \textbf{2023}.

\bibitem{Bouzenada2023NPB116288}
A.~Bouzenada, A.~Boumali, R.~L.~L. Vit{\'o}ria, F.~Ahmed, M.~Al-Raeei,
  \emph{Nuclear Physics B} \textbf{2023}, \emph{994}, 116288.

\bibitem{Bouzenada2024NPB116682}
A.~Bouzenada, A.~Boumali, F.~Ahmed, \emph{Nuclear Physics B} \textbf{2024},
  \emph{1007}, 116682.

\bibitem{Bouzenada2024TMP_SomRay}
A.~Bouzenada, A.~Boumali, R.~L.~L. Vit{\'o}ria, C.~Furtado, \emph{Theoretical
  and Mathematical Physics} \textbf{2024}, \emph{221}, 2193.

\bibitem{Puntigam_1997}
R.~A. Puntigam, H.~H. Soleng, \emph{Classical and Quantum Gravity}
  \textbf{1997}, \emph{14}, 1129--1149.

\bibitem{Ozdemir_2005}
N.~Ozdemir, \emph{International Journal of Modern Physics A} \textbf{2005},
  \emph{20}, 2821--2832.

\bibitem{Jusufi_2016}
K.~Jusufi, \emph{European Physical Journal C} \textbf{2016}, \emph{76}, 332.

\bibitem{Abramowitz_1964}
M.~Abramowitz, I.~A. Stegun, \emph{Handbook of Mathematical Functions with
  Formulas, Graphs, and Mathematical Tables}, \bblvol{}~55 \bblof{}
  \emph{National Bureau of Standards Applied Mathematics Series}, U.S.
  Government Printing Office, Washington, D.C., \textbf{1964}.

\bibitem{Andrews_2001}
G.~E. Andrews, R.~Askey, R.~Roy, \emph{Special Functions}, Cambridge University
  Press, Cambridge, \textbf{2001}.

\bibitem{Arfken_2012}
G.~B. Arfken, H.~J. Weber, F.~E. Harris, \emph{Mathematical Methods for
  Physicists: A Comprehensive Guide} \emph{7\bblth{} \bbledn{}}, Elsevier,
  Academic Press, Amsterdam, \textbf{2012}.

\bibitem{Fluegge_1994}
S.~Fl{\"u}gge, \emph{Practical Quantum Mechanics}, Springer, \textbf{1994}.

\bibitem{Silenko_2008}
A.~J. Silenko, \emph{Physical Review A} \textbf{2008}, \emph{77}, 012116.

\bibitem{Silenko_2008_TMP}
A.~J. Silenko, \emph{Theoretical and Mathematical Physics} \textbf{2008},
  \emph{156}, 1308.

\bibitem{Silenko_2013}
A.~J. Silenko, \emph{Physical Review D} \textbf{2013}, \emph{88}, 045004.

\bibitem{Stam1959}
A.~J. Stam, \emph{Information Control} \textbf{1959}, \emph{2}, 101.

\bibitem{Cramer1959}
H.~Cramer, \emph{Mathematical Methods Of Statisticsby}, Princeton University
  Press Princeton, \textbf{1959}.

\bibitem{BBM1975}
I.~Białynicki-Birula, J.~Mycielski, \emph{Commun.Math.Phys} \textbf{1975},
  \emph{44}, 129--132.

\bibitem{Weizsaecker1935}
C.~F.~v. Weizs{\"a}cker, \emph{Zeitschrift f{\"u}r Physik} \textbf{1935},
  \emph{96}, 431--458.

\bibitem{Moreira2025GodelEntropy}
A.~R.~P. Moreira, L.~Ma, A.~Bouzenada, F.~Ahmed, \emph{International Journal of
  Modern Physics A} \textbf{2025}, \emph{40}.

\bibitem{Yang2022CosmicStringQFI}
Y.~Yang, J.~Jing, Z.~Tian, \emph{The European Physical Journal C}
  \textbf{2022}, \emph{82}, 688.

\bibitem{Arvizu2024ConicalEntropy}
L.~M. Arvizu, E.~Casta{\~n}o, N.~Aquino, \emph{Physica Scripta} \textbf{2024},
  \emph{99}, 095270.

\bibitem{Lima2023QIE_AB}
F.~C.~E. Lima, A.~R.~P. Moreira, C.~A.~S. Almeida, C.~O. Edet, N.~Ali,
  \emph{Physica Scripta} \textbf{2023}, \emph{98}, 065111.

\bibitem{moshinsky1989}
M.~Moshinsky, A.~Szczepaniak, \emph{Journal of Physics A: Mathematical and
  General} \textbf{1989}, \emph{22}, L817.

\bibitem{Lamata2011}
L.~Lamata, J.~Casanova, R.~Gerritsma, C.~F. Roos, J.~J. Garcia-Ripoll,
  E.~Solana, \emph{New. J. Phys} \textbf{2011}, \emph{13}, 095003.

\bibitem{Blatt2012}
R.~Blatt, C.~F. Roos, \emph{Nat. Phys} \textbf{2012}, \emph{8}, 277--284.

\bibitem{FrancoVillafane2013DiracOscillator}
J.~A. Franco-Villafa\~{n}e, E.~Sadurn\'{\i}, S.~Barkhofen, U.~Kuhl,
  F.~Mortessagne, T.~H. Seligman, \emph{Physical Review Letters} \textbf{2013},
  \emph{111}, 170405.

\bibitem{Gerritsma2010DiracEquationIons}
R.~Gerritsma, G.~Kirchmair, F.~Z\"{a}hringer, E.~Solano, R.~Blatt, C.~F. Roos,
  \emph{Nature} \textbf{2010}, \emph{463}, 68--71.

\end{thebibliography}

\end{document}